\documentclass[
prc,%
10pt,%
final,%
notitlepage,%
oneside,%
twocolumn,%
nobibnotes,%
nofootinbib,
superscriptaddress,%
floatfix,%
showkeys,%
showpacs]%
{revtex4}
\usepackage{color}
\usepackage{amsfonts}
\usepackage{amsbsy}
\usepackage{mathrsfs}
\usepackage{graphicx}
\def\lsim{\mathrel{\rlap{
\lower4pt\hbox{\hskip-3pt$\sim$}}
    \raise1pt\hbox{$<$}}}     
\def\gsim{\mathrel{\rlap{
\lower4pt\hbox{\hskip-3pt$\sim$}}
    \raise1pt\hbox{$>$}}}     
\def\scr#1{\mbox{\scriptsize #1}}
\begin{document}
\title{Alternative Scenarios of Relativistic Heavy-Ion Collisions: 
I. Baryon Stopping} 
\author{Yu.B. Ivanov}\thanks{e-mail: Y.Ivanov@gsi.de}
\affiliation{Kurchatov Institute, 
Moscow RU-123182, Russia}
\begin{abstract}
Simulations of relativistic heavy-ion collisions within the three-fluid model
employing a purely hadronic equation of state (EoS) and two versions of the EoS
involving deconfinement transition are presented. The latter are 
an EoS with the first-order phase transition  
and that with a smooth crossover transition.
The model setup is described in detail. 
The analysis is performed   in a wide range of incident energies  
 2.7 GeV  $\le \sqrt{s_{NN}}\le$ 39 GeV in terms of the 
center-of-mass energy. 
Results on proton and net-proton rapidity distributions are reported. 
Comparison with available data indicate certain preference of the
crossover EoS.
It is found that predictions within  deconfinement-transition scenarios
exhibit a ``peak-dip-peak-dip'' irregularity in the 
incident energy dependence of the form of the net-proton rapidity distributions in central collisions. 
This irregularity is a signal of deconfinement onset occurring
in the hot and dense stage of the nuclear
collision.
\end{abstract}
\maketitle

\section{Introduction}

Over the last decade relativistic heavy ion physics has made tremendous progress
in understanding of the QCD phase diagram in the domain of high temperatures and 
low baryon density \cite{shuryak08}. However, a number of important questions still 
remain open. These are: {\em At which incident energy an onset of deconfinement happen? 
What is the order of the deconfinement transition at high baryon densities? 
Is there a critical end point in the phase diagram?} These questions form the main 
motivation for the currently running beam-energy-scan program \cite{RHIC-scan} at the 
Relativistic Heavy-Ion Collider (RHIC) at Brookhaven National Laboratory (BNL) and 
low-energy-scan program \cite{SPS-scan} at Super Proton Synchrotron (SPS)
of the European Organization for Nuclear Research (CERN), as well as newly constructed 
Facility for Antiproton and Ion Research (FAIR) in Darmstadt \cite{FAIR} and the
Nuclotron-based Ion Collider Facility (NICA) in Dubna \cite{NICA}.

This paper starts a series of papers, in which I hope 
to shed light on 
two first questions 
formulated above. In these papers I will report results of thorough simulations of 
relativistic heavy-ion collisions in the energy range from 2.7 GeV 
to 39 GeV (in terms of center-of-mass energy, $\sqrt{s_{NN}}$). This domain covers 
the energy range of the RHIC beam-energy-scan and SPS low-energy-scan programs, as well as 
energies of the future FAIR and NICA facilities and the Alternating Gradient
Synchrotron (AGS) at BNL. Though experiments at the AGS have been already stopped, 
experimental data taken at the AGS are still unique since they were neither updated nor repeated
in any newer measurements so far. The simulations were performed within a model of the three-fluid 
dynamics (3FD) \cite{3FD} employing three different equations of state (EoS): a purely hadronic EoS   
\cite{gasEOS} (hadr. EoS), which was used in the major part of the 3FD simulations so far 
\cite{3FD,3FD-GSI07,3FDflow,3FDpt,3FDv2}, and two versions of EoS involving the deconfinement 
transition \cite{Toneev06}. These two versions are an EoS with the first-order phase transition
and that with a smooth crossover transition. 
Since neither of EoS's includes a critical end point, these simulations do not touch the last 
question  formulated above.

This report is planned as a series of papers because it concerns  
a great number of  bulk observables (rapidity and transverse spectra, flow observables and multiplicities) 
for various species and a large number of incident energies, their comparison with available data, 
and also illustrations of global evolution of collisions within different scenarios (i.e. EoS's).  
Analysis of whole the set of observables  will be useful 
for revealing possible correlations in the energy evolution of these observables 
within different scenarios, which can be used as experimental indications of 
the deconfinement onset or its absence. 

It is reasonable to start this series with analysis of the baryon stopping because 
a degree of stopping of colliding nuclei
is one of the basic characteristics of 
the collision dynamics, which determines a part of the incident energy
of colliding nuclei 
deposited into produced fireball and hence into production of
secondary particles. The deposited energy in its turn determines the
nature (hadronic or quark-gluonic) of the produced fireball and
thereby its subsequent evolution. 
Therefore, a proper reproduction of the baryon stopping
is of prime importance for theoretical understanding of the dynamics 
of the nuclear collisions. I will argue that certain irregularity 
in the incident-energy dependence of the baryon stopping may indicate an onset of 
deconfinement. 

The paper is organized as follows. In the next section \ref{Model} 
a brief survey of the 3FD model is presented: basic ideas and choice of 
parameters relevant for the present simulations. 
Properties of the EoS's used in the present simulations are illustrated in sect. \ref{EOS}. 
Predictions of proton and net-proton rapidity distributions and their comparison 
with available experimental data are presented in sect. \ref{rapidity distributions}. 
Analysis of the form of these rapidity distributions and its evolution with incident 
energy rise is done in sect. \ref{Analysis}. 
In sect. \ref{Summary} a summury of results is formulated. 
Results on baryon stopping in simulations with deconfinement
transitions have been already briefly reported (without details) in Ref. \cite{Ivanov:2012bh}.

\section{3FD Model}\label{Model}

The 3FD model \cite{3FD} is a
straightforward extension of the 2-fluid model with radiation of
direct pions \cite{MRS88,gsi94,MRS91} and (2+1)-fluid model
\cite{Kat93,Brac97}. The above models were extend in such a
way that the created baryon-free fluid (which is called a
``fireball'' fluid, following the Frankfurt group) is treated
on equal footing with the baryon-rich ones. 
A certain formation time $\tau$ is allowed for the fireball fluid, during
which the matter of the fluid propagates without interactions. 
The formation time 
is  associated with a finite time of string formation. It is similarly 
incorporated in kinetic transport models such as UrQMD \cite{Bass98} 
and HSD \cite{Cassing99}.

Unlike the conventional hydrodynamics, where local instantaneous
stopping of projectile and target matter is assumed, a specific
feature of the 3FD is a finite stopping
power resulting in a counter-streaming regime of leading
baryon-rich matter. The basic idea of a 3-fluid approximation to
heavy-ion collisions \cite{MRS88,I87} is that at each space-time
point $x=(t,{\bf x})$ a generally nonequilibrium 
distribution function of baryon-rich
matter, $f_{\scr{br}}(x,p)$, can be represented as a sum of two
distinct contributions
\begin{eqnarray}
\label{t1} f_{\scr{br}}(x,p)=f_{\scr p}(x,p)+f_{\scr t}(x,p),
\end{eqnarray}
initially associated with constituent nucleons of the projectile
(p) and target (t) nuclei. In addition, newly produced particles,
populating the mid-rapidity region, are associated with a fireball
(f) fluid.
Therefore, the 3-fluid approximation is a minimal way to 
simulate the finite stopping power at high incident energies.

The above assumptions are implemented into the formulation of the 
3FD model as follows \cite{3FD}. There is a set of continuity equations 
(reflecting the baryon number conservation) 
   \begin{eqnarray}
   \label{eq8}
   \partial_{\mu} J_{\alpha}^{\mu} (x) &=& 0,
   \end{eqnarray}
for $\alpha=$p and t, where
$J_{\alpha}^{\mu}=n_{\alpha}u_{\alpha}^{\mu}$ is the baryon
current defined in terms of baryon density $n_{\alpha}$ and
 hydrodynamic 4-velocity $u_{\alpha}^{\mu}$ normalized as
$u_{\alpha\mu}u_{\alpha}^{\mu}=1$. Eq.~(\ref{eq8}) implies that
there is no baryon-charge exchange between p- and t-fluids, as
well as that the baryon current of the fireball fluid is
identically zero, $J_{\scr f}^{\mu}=0$. 
Equations of the energy--momentum exchange between fluids are formulated
in terms of 
energy--momentum tensors $T^{\mu\nu}_\alpha$ of the 
fluids
   \begin{eqnarray}
   \partial_{\mu} T^{\mu\nu}_{\scr p} (x) &=&
-F_{\scr p}^\nu (x) + F_{\scr{fp}}^\nu (x),
   \label{eq8p}
\\
   \partial_{\mu} T^{\mu\nu}_{\scr t} (x) &=&
-F_{\scr t}^\nu (x) + F_{\scr{ft}}^\nu (x),
   \label{eq8t}
\\
   \partial_{\mu} T^{\mu\nu}_{\scr f} (x) &=&
- F_{\scr{fp}}^\nu (x) - F_{\scr{ft}}^\nu (x)
\cr
&+&
\int d^4 x' \delta^4 \left(\vphantom{I^I_I} x - x' - U_F
(x')\tau\right)
\cr
&\times&
 \left[F_{\scr p}^\nu (x') + F_{\scr t}^\nu (x')\right],
   \label{eq8f}
   \end{eqnarray}
where the $F^\nu_\alpha$ are friction forces originating from
inter-fluid interactions. $F_{\scr p}^\nu$ and $F_{\scr t}^\nu$ in
Eqs.~(\ref{eq8p})--(\ref{eq8t}) describe energy--momentum loss of the 
baryon-rich fluids due to their mutual friction. A part of this
loss $|F_{\scr p}^\nu - F_{\scr t}^\nu|$ is transformed into
thermal excitation of these fluids, while another part $(F_{\scr
p}^\nu + F_{\scr t}^\nu)$ gives rise to particle production into
the fireball fluid (see Eq.~(\ref{eq8f})). $F_{\scr{fp}}^\nu$ and
$F_{\scr{ft}}^\nu$ are associated with friction of the fireball
fluid with the p- and t-fluids, respectively. 
Here $\tau$ is the formation time, and
   \begin{eqnarray}
   \label{eq14}
U^\nu_F (x')=
\frac{u_{\scr p}^{\nu}(x')+u_{\scr t}^{\nu}(x')}%
{|u_{\scr p}(x')+u_{\scr t}(x')|}
   \end{eqnarray}
is a free-propagating 4-velocity of the produced fireball 
matter.
Accordingly to Eq.~(\ref{eq8f}), 
this matter 
gets formed only 
after the time span $U_F^0\tau$ upon the
production, and in 
different space point ${\bf x}' - {\bf U}_F (x') \ \tau$, as
compared to the production point ${\bf x}'$.

The nucleon--nucleon cross sections at high energies are strongly
forward--backward peaked. This fact, which originally served as
justification for subdividing baryonic matter into target and
projectile fluids,
was used in \cite{IMS85} to estimate the friction forces, $F_{\scr
p}^\nu$ and $F_{\scr t}^\nu$, proceeding from only $NN$ elastic
scattering. Later these friction forces were calculated
\cite{Sat90} based on (both elastic and inelastic) experimental
inclusive proton--proton cross sections. In the present
calculations  the following form of the projectile--target
friction is used
 \begin{eqnarray}
 F_{\alpha}^\nu=\vartheta^2\rho^{\xi}_{\scr p} \rho^{\xi}_{\scr t}
\left[\left(u_{\alpha}^{\nu}-u_{\bar{\alpha}}^{\nu}\right)D_P+
\left(u_{\scr p}^{\nu}+u_{\scr t}^{\nu}\right)D_E\right],
\label{eq16}
\end{eqnarray}
$\alpha=$p or t, $\bar{\mbox{p}}=$t and  $\bar{\mbox{t}}=$p.
Here, $\rho^{\xi}_\alpha$  denotes a kind of "scalar
density" of the p- and t-fluids (see below),
\begin{eqnarray}
D_{P/E} = m_N \ V_{\scr{rel}}^{\scr{pt}} \ \sigma_{P/E}
 (s_{\scr{pt}}),
\label{eq17}
\end{eqnarray}
where $m_N$ is the nucleon mass, $s_{\scr{pt}}=m_N^2 \left(u_{\scr
p}^{\nu}+u_{\scr t}^{\nu}\right)^2$,  
$V_{\scr{rel}}^{\scr{pt}}=
[s_{\scr{pt}}(s_{\scr{pt}}-4m_N^2)]^{1/2}/2m_N^2$
is the mean relative velocity of the p- and t-fluids, and
$\sigma_{P/E}(s_{\scr{pt}})$ are determined in terms of
nucleon-nucleon cross sections integrated with certain weights
(see \cite{MRS88,MRS91,Sat90} for details):  
\begin{eqnarray}
\label{sigma_P} 
\hspace*{-5mm}
\sigma_{P}(s_{\scr{pt}}) &\!\!\!=\!\!\!& 
\int_{\theta_{\scr{cm}}<\pi/2} 
d\sigma_{NN\to NX} \left(1-\cos\theta_{\scr{cm}}
\frac{p_{\scr{out}}}{p_{\scr{in}}}\right),
\\
\label{sigma_E} 
\hspace*{-5mm}
\sigma_{E}(s_{\scr{pt}}) &\!\!\!=\!\!\!& 
\int_{\theta_{\scr{cm}}<\pi/2} 
d\sigma_{NN\to NX} \left(1-
\frac{E_{\scr{out}}}{E_{\scr{in}}}\right). 
\end{eqnarray}
Here the integration is restricted to the forward hemisphere 
($\theta_{\scr{cm}}<\pi/2$) of the 
center-of-mass scattering angles $\theta_{\scr{cm}}$, 
$p_{\scr{in}}=(s_{\scr{pt}}/4-m_N^2)^{1/2}$ and 
$E_{\scr{in}}=s_{\scr{pt}}^{1/2}/2$ are the in-coming momentum and
energy of the nucleon in the NN c.m. frame,
respectively, and $p_{\scr{out}}$ and $E_{\scr{out}}$ are the
corresponding out-coming quantities. $\sigma_{P}(s_{\scr{pt}})$
is nonzero at any physical
$s_{\scr{pt}}$, as it is seen from Eq. (\ref{sigma_P}). 
At the same time, the $\sigma_{E}(s_{\scr{pt}})$ quantity, which is
responsible for the fireball production, is
zero for $s_{\scr{pt}}$ below the inelastic threshold. 
The overall
$\vartheta^2$ factor in Eq. (\ref{eq16}) controls
unification of p- and t-fluid into a single one, 
when their relative velocity gets small enough 
(for details see \cite{3FD}).

The above friction (\ref{eq16}) is a certain extension of that
derived in \cite{Sat90}. The original derivation \cite{Sat90} was
performed under assumption that baryon-rich fluids consist of only
nucleons, and only proton--proton cross sections were used 
in (\ref{sigma_P}) and  (\ref{sigma_E}). 
The extension is required because the original derivation \cite{Sat90}
does not take into account:  
\\
(i) various mesonic and baryonic species produced in the collision
\\
(ii) possible multiparticle interactions which are quite probable
in the dense medium,
\\
(iii) possible medium modifications of cross sections and effective
masses, and 
\\
(iv) quark and gluon interactions, if deconfinement occurs. 
\\
In view of these uncertainties, it is reasonable to make
provision for tuning the above friction. For this purpose,  
 tuning factors $\xi (s_{pt})$ in the scalar densities
of the p- and t-fluids are introduced
\begin{eqnarray}
\label{ro_xi} 
\rho_{\alpha}^\xi(s_{pt}) &=&
\left(\rho_{\alpha}^{bar.}+\frac{2}{3}\rho_{\alpha}^{mes.}\right)\xi_h(s_{pt})
\cr
&+&
\frac{1}{3}\left(\rho_{\alpha}^{q}+\rho_{\alpha}^{g}\right)\xi_q(s_{pt}),
\end{eqnarray}
where 
$\rho_{\alpha}^{bar.}$, $\rho_{\alpha}^{mes.}$,
$\rho_{\alpha}^{q}$ and $\rho_{\alpha}^{g}$ are scalar densities
of all baryons, all mesons, quarks and gluons, respectively, 
defined in the conventional way. 
This quantities are supplied together with EoS. 
Factors like 2/3 and 1/3 in Eq.
(\ref{ro_xi}) take into account the assumed scaling of cross sections in
accordance with the naive valence-quark counting.
In view of above mentioned uncertainties of the estimated friction, 
in Eq. (\ref{ro_xi}) different tuning factors
are introduced for hadronic and quark-gluon phases: $\xi_h$ and
$\xi_q$,  respectively. 

The friction between baryon-rich fluids was fitted to reproduce
the stopping power observed in proton rapidity distributions for each EoS. 
The results, together with those for formation time $\tau$ and 
freeze-out energy density $\varepsilon_{\scr frz}$, are summarized as follows: 

\begin{itemize} 

\item
Hadronic EoS with incompressibility $K$ = 190 MeV  \cite{gasEOS} (hadr. EoS):  
\begin{eqnarray}
\label{xi} 
&&\xi_h^2 (s) =1+
\left[\ln\left(\frac{s^{1/2}}{2m_N}\right)\right]^{1/4}, \quad
\\
&&\tau = 2 \;\mbox{fm/c}, \quad \varepsilon_{\scr frz} = 0.4 \;\mbox{GeV/fm}^3. 
\label{tau} 
\end{eqnarray}
The $\xi_q$ factor is not applicable 
here because of the pure hadronic nature of the EoS.
The incompessibility $K$ = 190 MeV is also chosen on the condition 
of the best reproduction of available data. 

\item
EoS with the first-order deconfinement  transition \cite{Toneev06} (2-phase EoS): 
\begin{eqnarray}
\label{xi-2ph} 
\xi_h^2 (s) = 1, &\quad&
\xi_q^2 (s) = 60\;\frac{4 m_N^2}{s},
\\
\tau = 0.17 \;\mbox{fm/c}, &\quad& \varepsilon_{\scr frz} = 0.4 \;\mbox{GeV/fm}^3. 
\label{tau-2ph} 
\end{eqnarray}

\item
EoS with crossover deconfinement transition \cite{Toneev06} (crossover EoS): 
\begin{eqnarray}
\label{xi-cr} 
\xi_h^2 (s) = 1, &\quad& 
\xi_q^2 (s) = 200\;\frac{4 m_N^2}{s},
\\
\tau = 0.17 \;\mbox{fm/c}, &\quad& \varepsilon_{\scr frz} = 0.4 \;\mbox{GeV/fm}^3. 
\label{tau-cr} 
\end{eqnarray}

\end{itemize}

Within hadronic scenario (hadr. EoS) the  friction 
has to be enhanced in order 
to reproduce the baryon stopping at high energies, $E_{lab} \geq 10A$ GeV. 
Though such a enhancement is admissible in view of above mentioned uncertainties, 
the value of the enhancement looks suspiciously high. 
Indeed, at $\sqrt{s_{NN}}= 17.3$ GeV, i.e. at the top SPS energy, 
$\xi_h^2=$ 2.2. 

At scenarios with deconfinement transitions there is no need to modify the hadronic friction. 
This can be considered as an indirect argument in favor of such scenarios. 
At the same time, the quark-gluon modification factor $\xi_q^2$ decreases with 
the energy rise, which is in agreement with our expectations that the 
quark-gluon friction should get weaker at high energies because of approaching to 
regime of the asymptotic freedom.

Freeze-out  was performed accordingly to the procedure described in \cite{3FD} 
and in more detail in \cite{71,74}. The baryon stopping turns out to be only 
moderately sensitive to the freeze-out energy density $\varepsilon_{\scr frz}$. 
The freeze-out energy density was chosen mostly on the condition of the best reproduction 
of secondary particles yields.

The formation time $\tau$ also affects the baryon stopping, especially at high incident
energies (top SPS and higher ones), when the fireball fluid is well developed. 
In fact, $\tau$ reduces the effect of the friction
between the fireball and baryon-rich fluids. The larger $\tau$ is, the later this friction 
starts to act and hence the weaker effect is produced by this friction. Therefore, 
the fitted value of $\tau$ is essentially related to the strength of the 
fireball--baryon-rich friction.


There are other full-scale (i.e. (3+1)-dimensional) approaches to modeling  
nuclear collisions, which take into account the deconfinement trasition. 
These have their advantages, as well as disadvantages 
as compared to the 3FD model. The conventional hydrodynamical model of Refs. 
\cite{Merdeev:2011bz,Mishustin:2010sd} does such simulations in a very 
similar way but without taking into account incomplete stopping of 
colliding nuclei at the initial stage of the reaction. Therefore, such kind of 
simulations are justified only at moderately high energies. Another class of fluid 
models uses (hadronic) kinetic codes to ``cook'' the initial fireball which is subsequently 
considered within the hydro simulation with possible deconfinement transitions 
\cite{Bleicher08,Bleicher09,Hama:2004rr,Nonaka:2012qw}. Such approaches 
disregard effects of deconfinement transitions at the stage of inter-penetration of 
colliding nuclei, and hence cannot be used for analysis of the baryon stopping, 
which is the prime goal of this article. Kinetic models with a deconfinement transition, 
i.e. A Multi-Phase Transport model \cite{Ko08} and a more consistent model of 
Parton-Hadron-String Dynamics \cite{Bratk09}, overcome this problem. Moreover, 
they avoid the problem of freeze-out inherent in hydrodynamic models. 
However, kinetic models are able to treat 
only a crossover  transition. The first-order transition remains beyond the 
scope of kinetics. Contrary to other hydrodynamic approaches, 
the 3FD model  can treat a deconfinement transition at the 
initial stage of the collision with due account of incomplete stopping of 
colliding nuclei, though in essentially rougher approximation than 
that in kinetic models. At the same time the 3FD model is able to work with 
the first-order transition unlike kinetic models.

In the present run of computations, higher incident 
energies were reached as compared with previous runs. 
This became possible because of implementation of 
an adaptive grid in the code. The size of the cell is made gradually larger 
with  the expansion of the system proceeded. Thus, when the system 
occupies the larger space, the code does not require a larger number of 
cells and, hence, a higher RAM memory. The adaptive grid does not make the 
accuracy worse since spatial distributions become smoother at the expansion,  
that relaxes requirements on the grid step. The adaptive grid made possible 
computations up to 62.4 GeV incident energy in terms of the c.m. nucleon-nucleon 
energy, i.e. $\sqrt{s_{NN}}$. However, results for the top energy of 
62.4 GeV 
are still not quite accurate, since an accurate computation requires 
unreasonably high memory and CPU time. This should be kept in mind when 
results for this energy are displayed.

\section{Equations of State} 
\label{EOS}

\begin{figure}[bth]
\includegraphics[width=7.0cm]{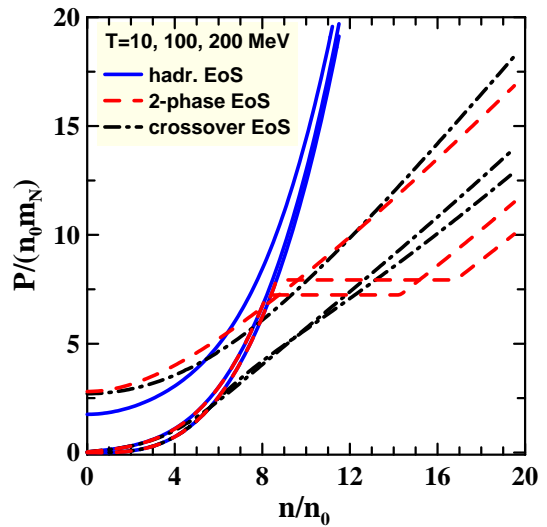}
 \caption{(Color online)
Pressure scaled by the product of normal nuclear density ($n_0=$ 0.15 fm$^{-3}$) and 
nucleon mass ($m_N$) versus baryon density scaled by the normal nuclear density
for three considered equations of state. Results are presented for three different
temperatures $T=$ 10, 100 and 200 MeV (bottom-up for corresponding curves).  
} 
\label{fig3.1}
\end{figure}

Figure \ref{fig3.1} illustrates differences between three considered EoS's. 
The deconfinement transition makes a EoS softer at high densities. 
The 2-phase EoS is based on the Gibbs construction, taking into account simultaneous conservation 
baryon and strange charges. However, the displayed result looks very similar to 
the Maxwell construction, corresponding to conservation of only baryon charge, 
with the only difference that the plateau is slightly tilted, which is practically invisible. 
This invisible slope of the plateau results from plotting 
the pressure at the additional condition of strange density being equal to zero
rather than at constant strange chemical potential. 
Application the Gibbs construction
in hydrodynamical simulations silently assumes that the inter-phase equilibration
in the mixed-phase region is faster than the hydrodynamical evolution.

%

%
\begin{figure}[tbh]
\includegraphics[width=7.0cm]{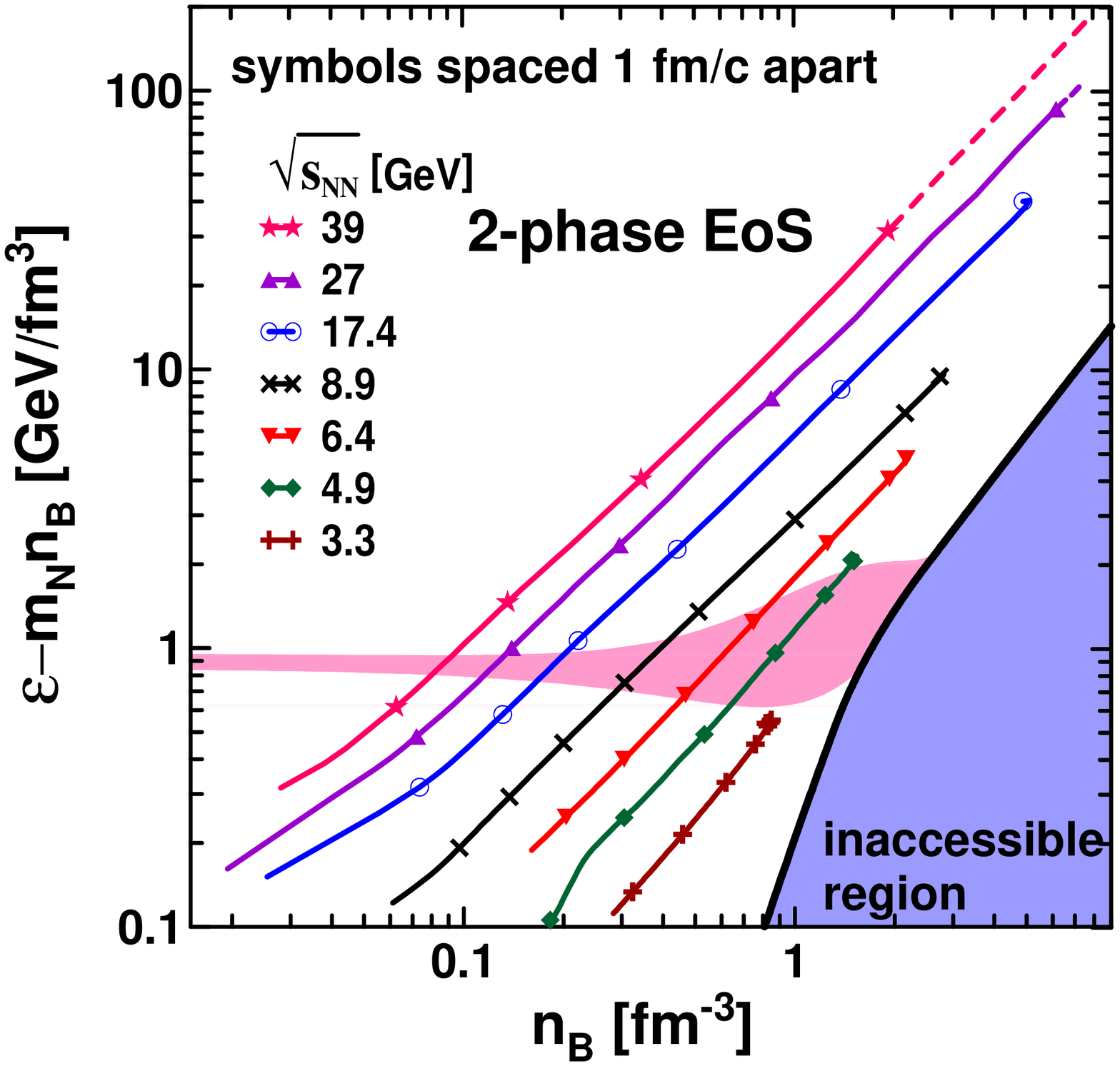}
\includegraphics[width=7.0cm]{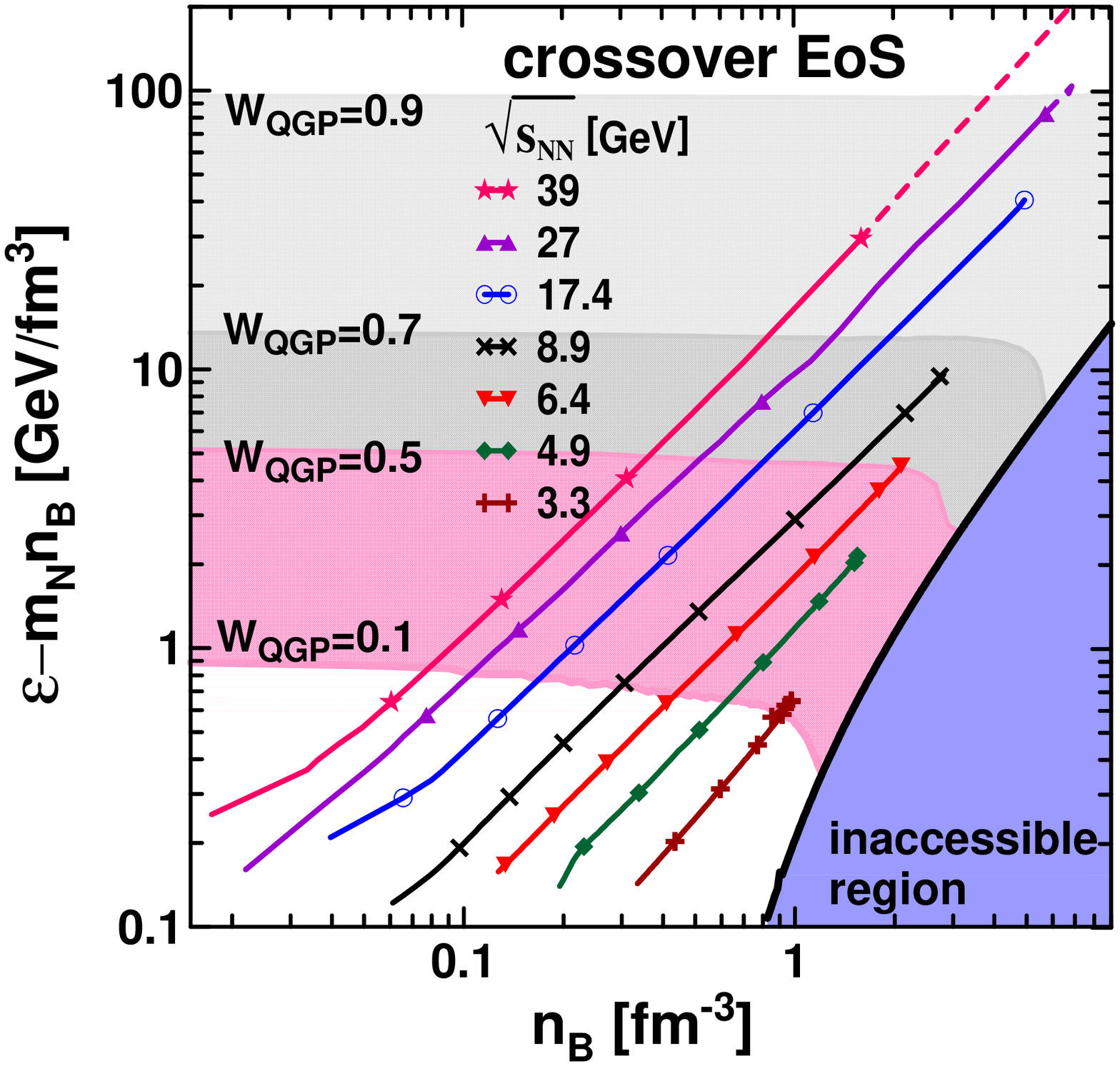}
 \caption{(Color online)
Dynamical trajectories of the matter in the central box of the
colliding nuclei  
(4fm$\times$4fm$\times \gamma_{cm}$4fm), where $\gamma_{cm}$ is the Lorentz
factor associated with the initial nuclear motion in the c.m. frame, 
for central collisions of Au+Au at 
$\sqrt{s_{NN}}=$ 3.3, 4.9, 27 and 39 GeV ($b=$ 2 fm), 
and Pb+Pb at $\sqrt{s_{NN}}=$ 6.4, 8.9 and 17.4 ($b=$ 2.4 fm). 
The trajectories are plotted in terms
of baryon density ($n_B$) and 
the energy density minus $n_B$ multiplied by the nucleon mass 
($\varepsilon - m_N n_B$). 
Only expansion stages of the
evolution are displayed.  
Symbols on the trajectories indicate the time rate of the evolution:
time span between marks is 1 fm/c. 
For the 2-phase EoS (upper panel) 
the shadowed ``mixed phase'' region is located between the borders, 
where the QGP phase start to raise ($W_{QGP}=$ 0) and becomes completely 
formed ($W_{QGP}=$ 1). 
For the crossover EoS  (lower panel) the displayed borders correspond to
values of the QGP fraction $W_{QGP}=$ 0.1, 0.5, 0.7 and 0.9.
Inaccessible region is restricted by $\varepsilon(n_B,T=0)-m_N n_B$ from above. 
}
\label{fig3.2}
\end{figure}

\begin{figure}[tbh]
\includegraphics[width=7.0cm]{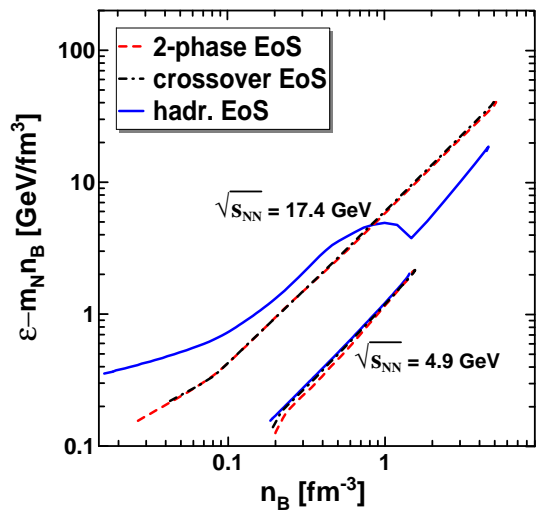}
 \caption{(Color online)
The same as in Fig. \ref{fig3.2} but for trajectories for all three 
different EoS's within the same frame. 
}
\label{fig3.3}
\end{figure}

The 2-phase and crossover EoS's still differ even at very high densities. 
The latter means that the crossover transition constructed in Ref. \cite{Toneev06}
is very smooth. The hadronic fraction survives up to very high densities. In particular, 
this is seen from Fig. \ref{fig3.2}: the fraction of the quark-gluon phase 
($W_{QGP}$) reaches value of 0.5 only at very high energy densities. 
In this respect, this version of the crossover EoS certainly contradicts results of the 
lattice QCD calculations, where a fast crossover, at least at zero chemical potential, 
was found \cite{Aoki:2006we}. 
Therefore, a true EoS is somewhere in between  the crossover and 2-phase EoS's 
of Ref. \cite{Toneev06}.

Figure \ref{fig3.2} demonstrates that the onset of the deconfinement transition
in the calculations happens at top-AGS--low-SPS energies. 
Similarly to that it has been done in \cite{Randrup07}, 
the figure displays dynamical 
trajectories of the matter in the central box placed around the
origin ${\bf r}=(0,0,0)$ in the frame of equal velocities of
colliding nuclei:  $|x|\leq$ 2 fm,  $|y|\leq$ 2 fm and $|z|\leq$
$\gamma_{cm}$ 2 fm, where $\gamma_{cm}$ is Lorentz
factor associated with the initial nuclear motion in the c.m. frame.  
Initially, the colliding nuclei are placed symmetrically with respect
to the origin ${\bf r}=(0,0,0)$, $z$ is the direction of the beam.
At a given density $n_B$, the zero-temperature
compressional energy, $\varepsilon(n_B,T=0)$, provides a lower bound on
the energy density $\varepsilon$, so the accessible region is correspondingly
limited. 
In the case of the crossover EoS only the region of the mixed phase between 
$W_{QGP}=$ 0.1 and $W_{QGP}=$ 0.5 is displayed, since in fact the mixed phase
occupies the whole ($\varepsilon$-$n_B$) region. 
The $\varepsilon$-$n_B$ representation
is chosen because these densities 
are dynamical quantities and, therefore, are suitable to compare
calculations with different EoS's.

\begin{figure*}[bt]
\hspace*{-14mm}\includegraphics[width=9.2cm]{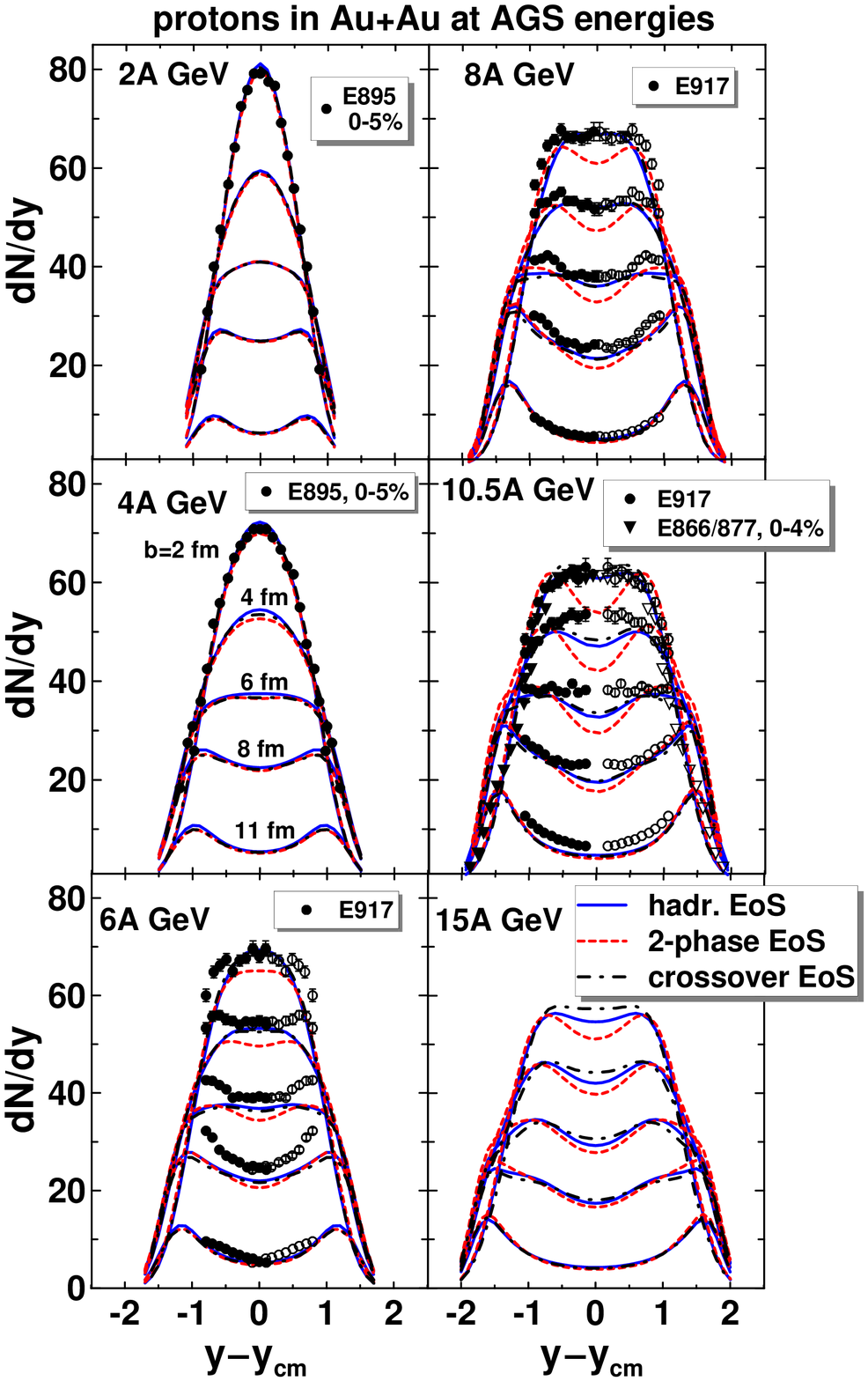}
\includegraphics[width=8.3cm]{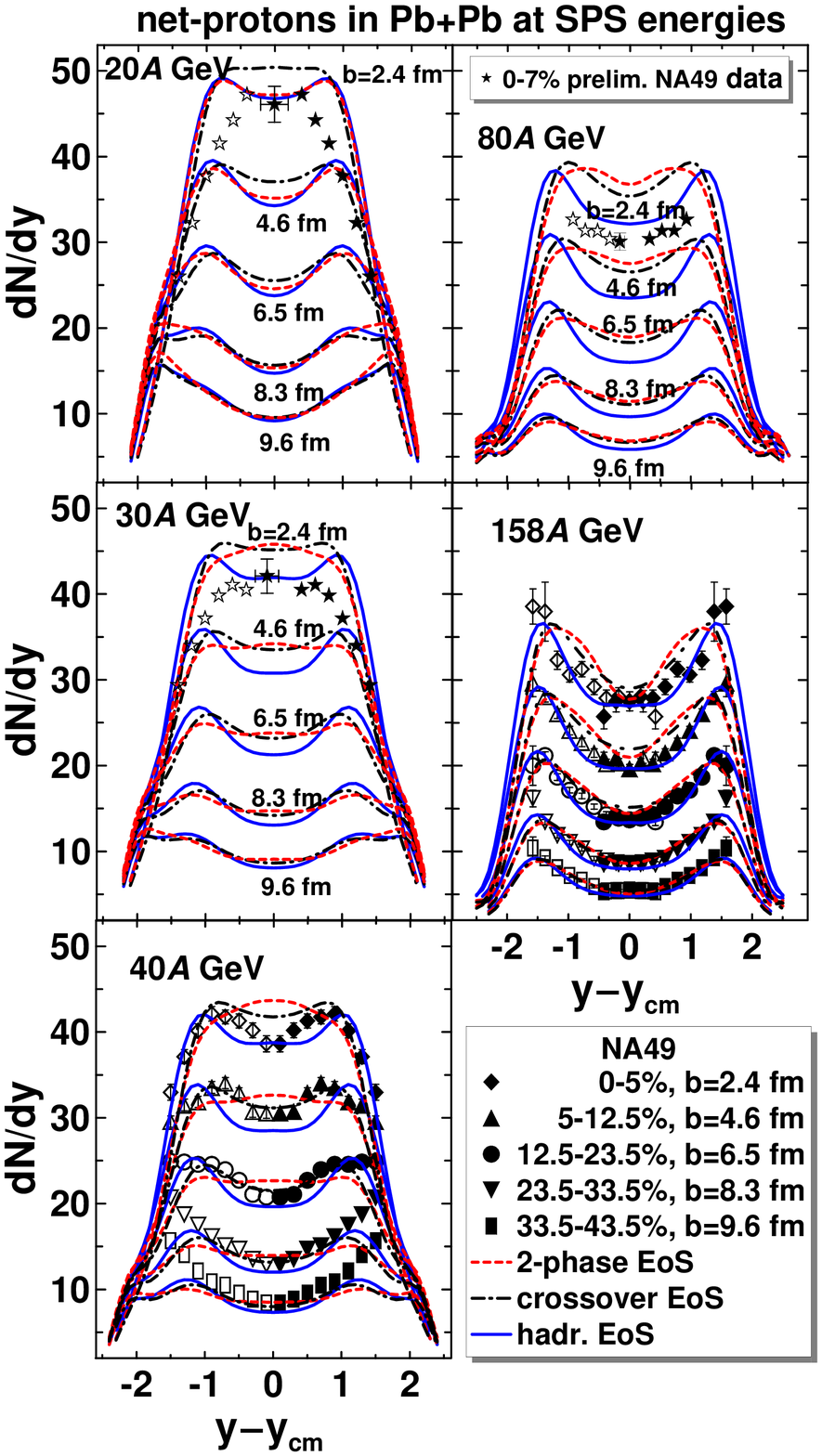}\hspace*{-14mm}
 \caption{(Color online)
Rapidity spectra of protons (for AGS energies, left block of panels)
and net-protons $(p-\bar p)$ (for SPS energies, right block of panels) from  
collisions of Au+Au (AGS) and Pb+Pb (SPS) calculated within three considered scenarios. 
Experimental data are from
collaborations E895 \cite{E895}, E877 \cite{E877},
E917 \cite{E917}, E866 \cite{E866}, and 
NA49 \cite{NA49-1,NA49-04,NA49-06,NA49-07,NA49-09}. 
The percentage shows the fraction of the total reaction cross section, 
corresponding to experimental selection of events. For the E917 data  
\cite{E917} these are 0-5\%, 5-12\%, 12-23\%, 23-39\% and 39-81\%.
Feedback of weak decays into $p$ and $\bar p$ yields is disregarded. 
} 
\label{fig4.1}
\end{figure*}

Only expansion stages of the evolution are displayed, where the matter
in the box is already thermally equilibrated, as a rule. The exceptions are 
central collisions at $\sqrt{s_{NN}}=$ 27 and 39 GeV, in which  
the matter in the box is not still thermalized in the beginning of the expansion stage. 
This non-equilibrium stage of the expansion is displayed by dashed lines 
in Fig. \ref{fig3.2}. 
The criterion of the thermalization is equality of longitudinal  and transverse
pressures in the box with the accuracy better than 10\%. 
Evolution proceeds from the top point of the trajectory downwards.
Symbols mark the time intervals along the trajectory. 
Subtraction of the $m_N n_B$ term is taken for the sake of suitable 
representation of the plot. 
The size of the box was chosen 
to be large enough that the amount of matter in it can be
representative to conclude on the onset of deconfinement 
and to be small enough to consider the matter in it as a homogeneous
medium. Nevertheless, the matter in the box still amounts to a minor part
of the total matter of colliding nuclei.  
Therefore, only the minor part of the total matter undergoes  the
deconfinement transition at 10$A$ GeV energy. 

As seen, the deconfinement transition starts at the top AGS energies in both cases. 
It gets practically completed at low SPS energies in the case of the   
2-phase EoS. In the crossover scenario it lasts till very high incident energies.

The trajectories for different  EoS's are very similar at lower 
 incident energies, as it seen from Fig. \ref{fig3.3}. At higher 
 energies trajectories for deconfinement scenarios remain very similar, 
 while the hadronic-EoS trajectories differ from those mentioned above and 
 exhibit a peculiar behavior. It happens because of a long 
 (as compared with the interpenetration time of colliding nuclei) 
 formation time of the fireball fluid, see (\ref{tau}). 
At the first stage the expansion proceeds when the fireball fluid has not been formed yet. 
Then the formation starts and the energy density (but not the baryon density) 
even slightly rises. When the formation is practically completed, the 
trajectory returns to its normal evolution -- downward in energy and baryon densities.

\section{Proton and Net-Proton rapidity distributions}
\label{rapidity distributions}

\begin{figure*}[htb]
\includegraphics[width=12.90cm]{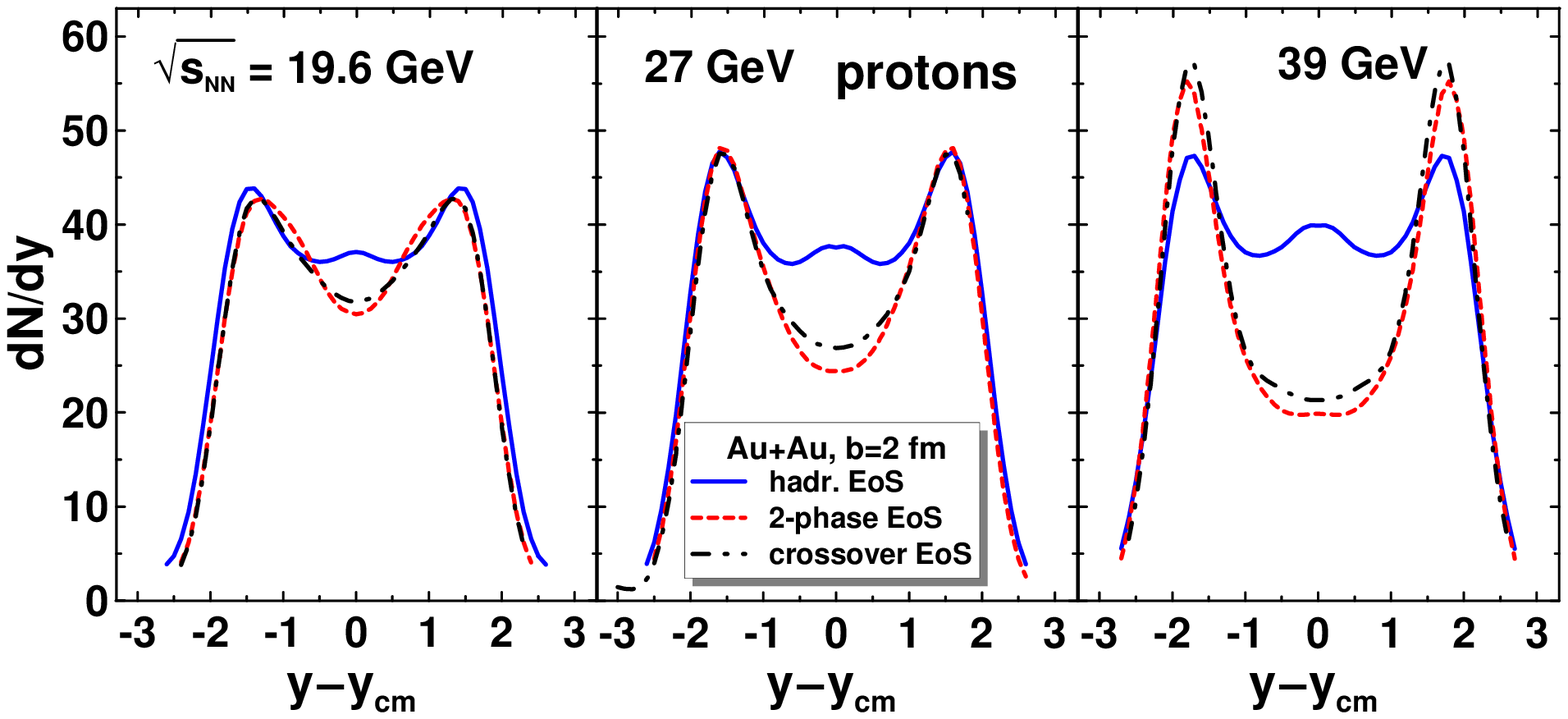}\\
\includegraphics[width=12.90cm]{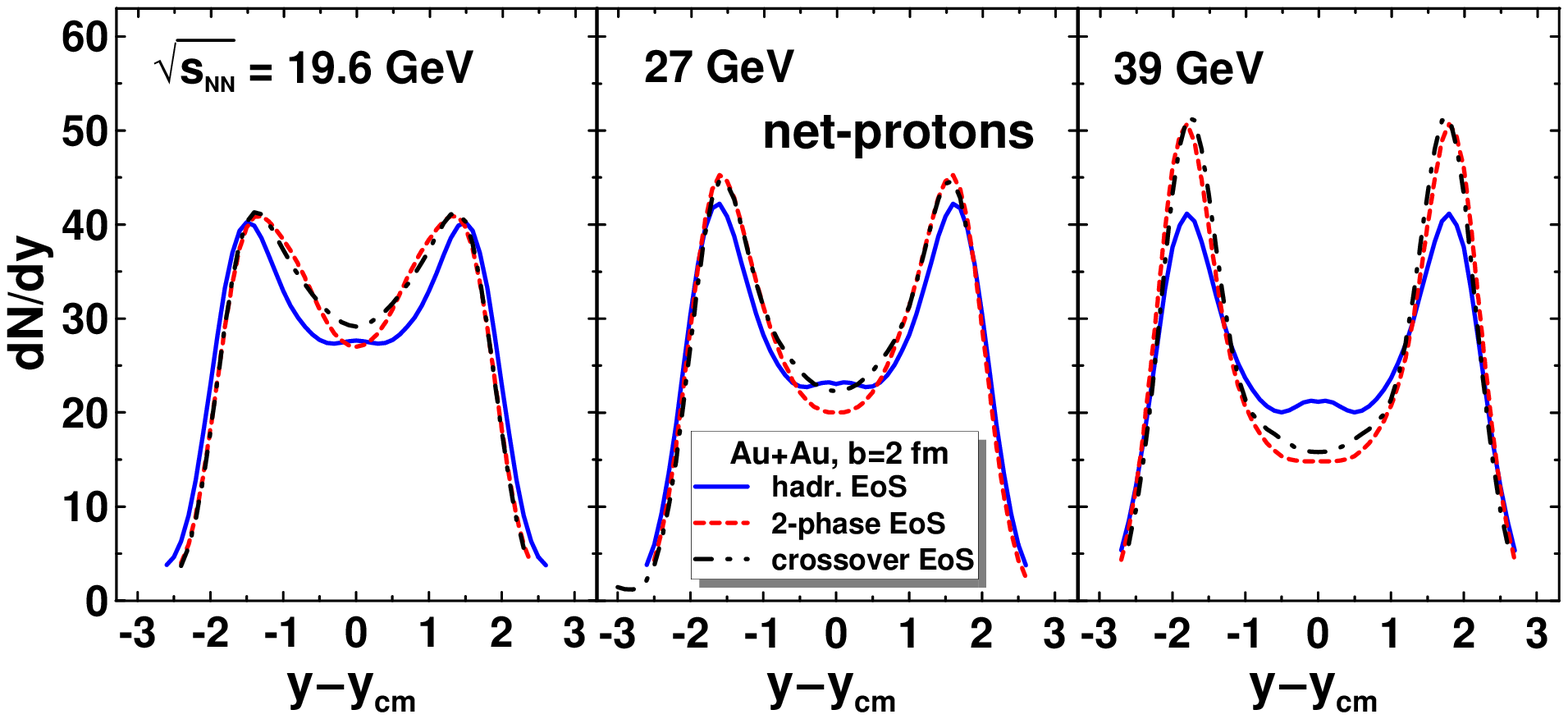}
 \caption{(Color online)
Rapidity spectra of protons (upper panel)
and net-protons (lower panel) from central 
collisions of Au+Au ($b=$ 2 fm) at low RHIC energies. 
Feedback of weak decays into $p$ and $\bar p$ yields is taken into account. 
} 
\label{fig4.2}
\end{figure*}

A direct measure of the baryon stopping is the
net-baryon (i.e. baryons-minus-antibarions) rapidity distribution. However, since experimental
information on neutrons is unavailable, we have to rely on net-proton (i.e. proton-minus-antiproton) data. 
Presently there exist experimental data on proton (or net-proton) rapidity spectra at 
AGS \cite{E895,E877,E917,E866} and 
SPS \cite{NA49-1,NA49-04,NA49-06,NA49-07,NA49-09} energies. 
These data were analyzed within various models  
\cite{3FD,3FD-GSI07,Bratk09,Merdeev:2011bz,Mishustin:2010sd,Bleicher09,Bratk04,WBCS03,Bratk02,Larionov07,Larionov05}. 
The most extensive analysis has been done in 
\cite{Bratk02,3FD}. 
Here I would like to repeat this analysis. 
The motivation is to perform simulations with different EoS's 
within the same dynamical model, i.e. the 3FD model, in order to 
reveal differences produced by different scenarios.

Figure \ref{fig4.1} presents calculated rapidity distributions of protons (for AGS energies) 
and net-protons (for SPS energies) and their comparison with available data. Notice that 
difference between protons and net-protons is negligible at the AGS energies. 
At the top AGS energy of 10$A$ GeV their difference is 0.03\% at the midrapidity,  
see compilation of experimental data in Ref. \cite{Andronic:2005yp}. 
Contribution of weak decays of strange hyperons into proton yield was disregarded
in accordance with mesurement conditions of the NA49 collaboration. 
At the AGS energies the contribution of weak decays is negligible. 
Correspondence between the fraction of the total reaction cross section related to a data set 
and a mean value of  the impact parameter was read off from the paper \cite{Alt:2003ab}
in case of NA49 data. 
For Au+Au collisions it was approximately estimated proceeding from geometrical considerations.

\begin{figure}[tbh]
\includegraphics[width=6.1cm]{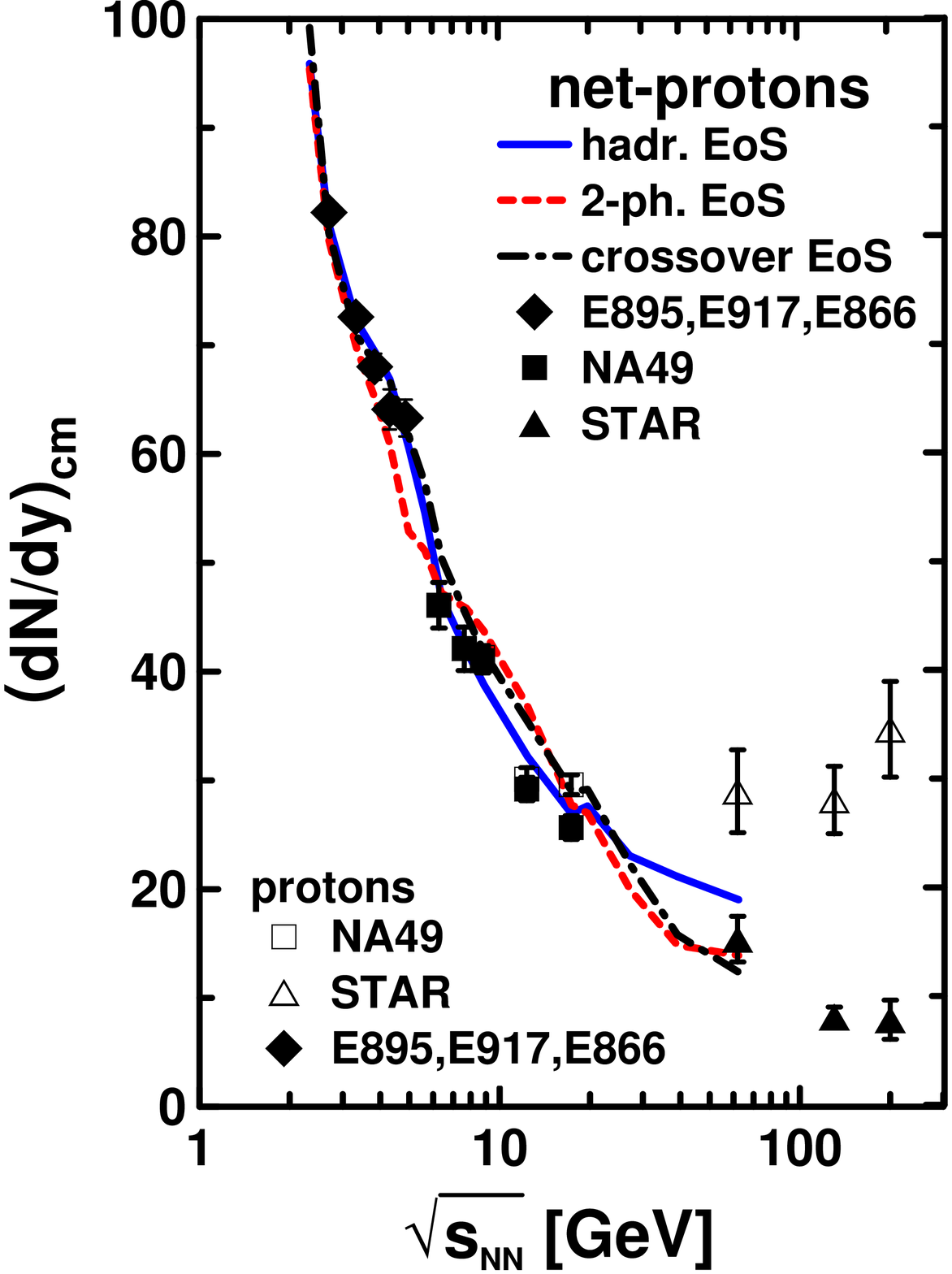}
\includegraphics[width=6.1cm]{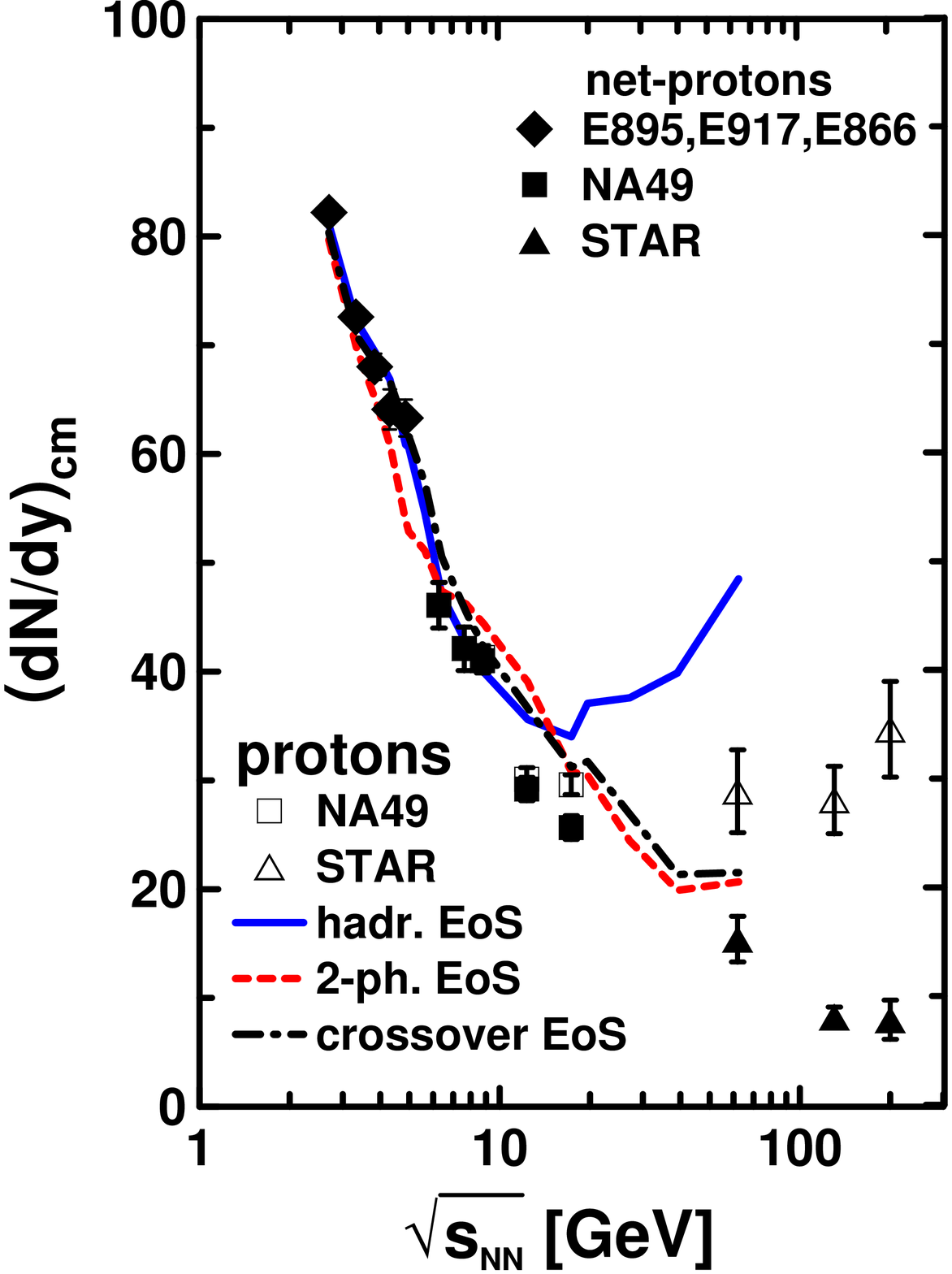}
 \caption{(Color online)
Midrapidity  value 
of the net-proton (upper panel) and proton (lower panel) rapidity
   spectrum from central Au+Au (at AGS and RHIC energies, $b=$ 2 fm) 
and Pb+Pb (at SPS energies, $b=$ 2.4 fm) collisions    
   as a function of the incident center-of-mass energy
for three considered EoS's. 
Experimental data are from
collaborations E895 \cite{E895}, E877 \cite{E877},
E917 \cite{E917}, E866 \cite{E866},  
NA49 \cite{NA49-1,NA49-04,NA49-06,NA49-07,NA49-09} and 
STAR  \cite{STAR09}. 
}  
\label{fig4c}
\end{figure}

As seen from Fig. \ref{fig4.1}, at lower AGS energies all EoS's predict the same results, 
since at these energies only hadronic parts of all EoS's are relevant.  
Results of the 
2-phase EoS start to differ from those of the hadr. and crossover EoS's beginning
from 6$A$ GeV and first in central collisions.  
At higher energies this difference extends to more peripheral collisions. 
Unlike other scenarios, the 2-ph.-EoS distributions exhibit a dip at midrapidity even 
in central collisions. 
This dip contradicts the available experimental data and is very robust: 
variation of the model parameters  (\ref{xi-2ph}) and (\ref{tau-2ph}) in a 
wide range does not remove this dip. Therefore, it is a direct consequence 
of the onset of the first-order phase transition, which starts precisely at these 
energies in the 2-phase scenario, see Fig. \ref{fig3.2}. 
Calculations within one-fluid (i.e. conventional hydrodynamics) 
\cite{Merdeev:2011bz,Mishustin:2010sd} confirm this conclusion. This dip survives 
even in one-fluid calculations involving the 1st-order phase transition in spite of 
immediate baryon stopping inherent in the one-fluid model. Notice that one-fluid calculations 
without deconfinement transition manifest a ``normal'' (for the one-fluid model) result, i.e. 
no dip. 
Accordingly to analysis of Ref. \cite{Merdeev:2011bz} this dip is a consequence of 
 larger pressure gradients in the longitudinal direction developed in the deconfinement-transition scenario.
In 3FD calculations,  this dip transforms into midrapidity peak at higher energies (30$A$ GeV and  40$A$ GeV). 
With further energy rise ($E_{lab}>$ 40$A$ GeV) the midrapidity peak again turns into a dip, 
see also  Fig. \ref{fig4.2}. 
The latter dip is already a normal behavior which results from incomplete stopping of baryons
and takes place at arbitrary high energies.

As has been already mentioned, the behavior ``peak-dip-peak-dip'' in central collisions within
the 2-phase-EoS scenario is very robust 
with respect to variation of the model parameters  (\ref{xi-2ph}) and (\ref{tau-2ph}) in a 
wide range. It certainly disagrees with data at 8$A$ GeV, 10$A$ GeV and  40$A$ GeV energies. 
It also disagrees with data at 20$A$ GeV and  30$A$ GeV. 
However, the latter data have preliminary status, and hence it is too early to draw any conclusions 
from comparison with them.  This behavior is in contrast with that for the hadronic-EoS scenario, 
where the form of distribution in central collisions gradually evolves from peak at the 
midrspidity to a dip. The case of the crossover EoS is intermediate. One could conclude 
in favor of a weak wiggle, since the distributions at 10$A$ and 15$A$ GeV exhibit a shallow dip while at 
20$A$ GeV looks like a plateau. 

Beginning from 158$A$ GeV to higher incident energies (see Fig. \ref{fig4.2}) predictions 
of different scenarios for the net-proton distributions remain quite similar, 
at the expense of the substantial enhancement of the hadronic friction in the case 
of the hadronic EoS, see Eq. (\ref{xi}). At the same time difference of proton spectra 
increase with the energy rise.  In calculations for energies above 158$A$ GeV 
contribution of weak decays into proton and net-proton yields were taken in to account 
in accordance of experimental procedure of STAR and PHENIX collaborations.

Comparison with available data indicate certain preference of the crossover EoS, 
though  the crossover scenario does not perfectly reproduce the data either.  
Predictions of different scenarios for net-protons diverge to the largest extent
in the energy region  8$A$  GeV  $\leq E_{lab} \leq$ 40$A$  GeV. 
Unfortunately data at 20$A$ and 30$A$  GeV still have preliminary status and 
disagree with any considered scenario. 
Updated experimental results at energies 20$A$ and 30$A$  GeV are badly needed 
to pin down the preferable EoS and to check a trend of the  ``peak-dip-peak-dip'' 
irregularity in the net-proton rapidity distributions.

Preference of the  deconfinement-transition scenarios is seen at incident 
energies above the top SPS one, 
see Fig. \ref{fig4c}, 
where midrapidity values  of the net-proton  and proton  rapidity
spectra from central collisions of Au+Au (at AGS and RHIC energies) 
and Pb+Pb (at SPS energies) are plotted as functions of the incident 
center-of-mass energy. 
In Fig. \ref{fig4c} the midrapidity  values 
are displayed in a wider energy range. 
I would like to remind that results for top 
calculated energy of $\sqrt{s_{NN}}=$ 62.4 GeV are very approximate, 
since a more accurate computation requires unreasonably high memory and CPU time. 
As seen, a visible difference between net-protons and proton data, 
as well as between predictions of hadronic EoS and EoS's with deconfinement transitions 
starts only at RHIC energies. At these energies the hadronic scenario 
certainly overestimates  the proton  midrapidity density.

\section{Analysis of ``peak-dip-peak-dip'' irregularity}
\label{Analysis}

\begin{figure*}[tbh]
\includegraphics[width=11.5cm]{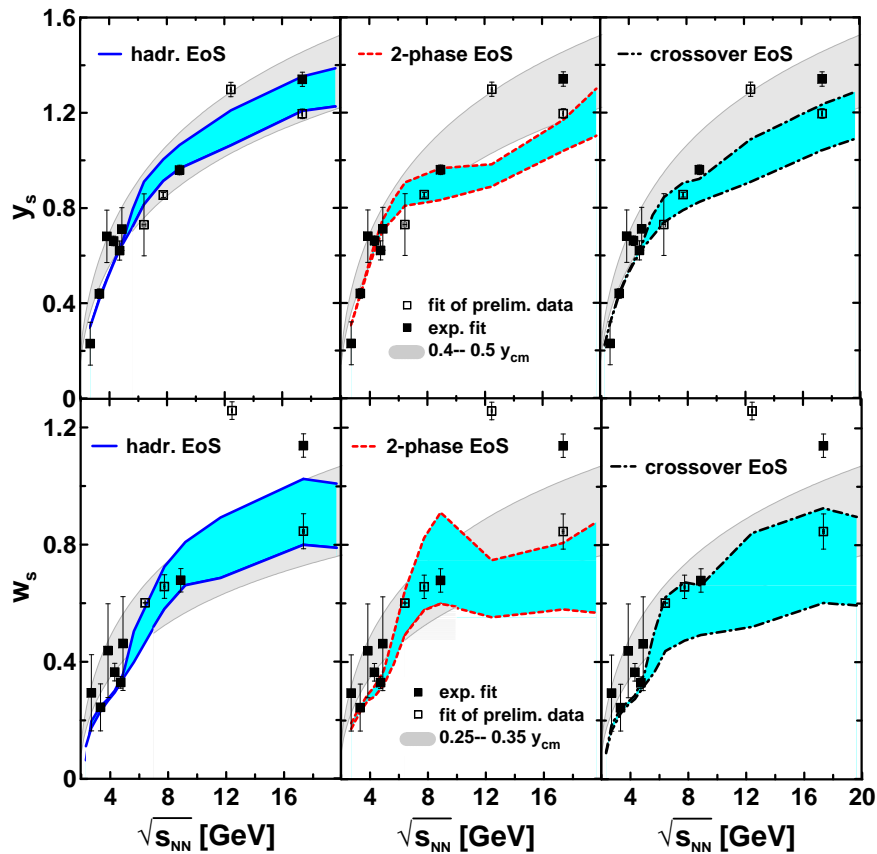}
 \caption{(Color online)
Parameters of fit (\ref{2-sources-fit}), $y_s$ (upper raw of panels) and $w_s$ 
(lower raw of panels), deduced from analysis of experimental data 
of net-proton rapidity distributions (points with error bars), as well as from that of 
results of 3FD simulations with different EoS's (shaded areas between respective lines). 
Fits of preliminary experimental data are displayed by open symbols, while those for 
confirmed data - by filled symbols. Lines, restricting shaded areas for different EoS's, 
are obtained by fits within different rapidity ranges: 
upper curves in the range of $|y-y_{cm}|/y_{cm}<0.5$, lower curves in $|y-y_{cm}|/y_{cm}<0.7$. 
Grey bands in all the panels indicate the areas: between $0.25\, y_{cm}$ and $0.35\, y_{cm}$ for 
$w_s$ (lower raw of panels), and $0.4\, y_{cm}$ and $0.5\, y_{cm}$ for $y_s$ (upper raw of panels). 
}  
\label{fig4aa}
\end{figure*}
\begin{figure*}[tbh]
\includegraphics[width=13.1cm]{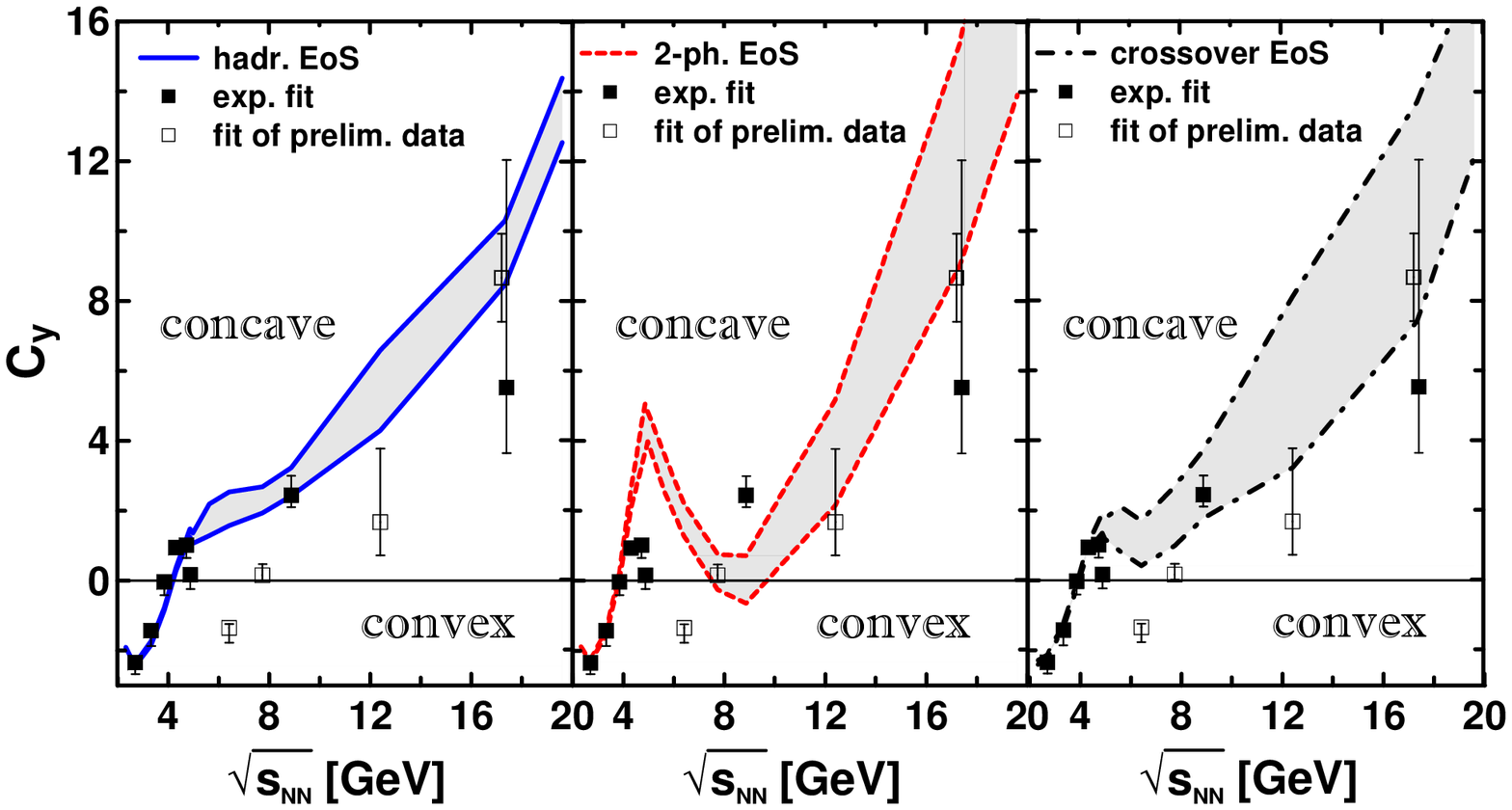}
 \caption{(Color online)
Midrapidity  reduced curvature  
   [see. Eq. (\ref{Cy})] 
of the net-proton rapidity
   spectrum as a function of the incident center-of-mass energy
 of colliding nuclei as deduced from experimental data and predicted
 by 3FD calculations with  different EoS's. 
Upper borders of the shaded areas correspond to fits confined in the region of $|y-y_{cm}|/y_{cm}<0.7$,  
lower borders -- $|y-y_{cm}|/y_{cm}<0.5$. 
}  
\label{fig4a}
\end{figure*}
%

%

Preliminary results of the above-discussed 
``peak-dip-peak-dip'' irregularity
 have been already reported in Refs. \cite{Ivanov:2010cu,Ivanov:2011cb}. 
There the friction forces for the 2-phase and crossover scenarios were poorly tuned and hence 
the corresponding simulations poorly reproduced available experimental data. 
Therefore, conclusions were  
based on a certain trend of the results of simulations. 
Here I present calculations with thoroughly tuned friction forces in the quark-gluon phase, 
which made it possible to reasonably 
(and often better than in the hadronic scenario)
reproduce a great number of observables in 
a wider (than before \cite{Ivanov:2010cu,Ivanov:2011cb}) 
incident energy range 2.7 GeV  $\le \sqrt{s_{NN}}\le$ 39 GeV.

%
In order to quantify the  
``peak-dip-peak-dip'' irregularity, it is useful to make use of the method 
proposed in Refs. \cite{Ivanov:2012bh,Ivanov:2010cu,Ivanov:2011cb}. For this purpose
the data on net-proton rapidity distributions are fitted  by a simple formula 
\begin{eqnarray}
\label{2-sources-fit} 
\frac{dN}{dy}&=&  
a \left(
\exp\left\{ -(1/w_s)  \cosh(y-y_{cm}-y_s) \right\}
\right.
\cr
&+&
\left.
\exp\left\{-(1/w_s)  \cosh(y-y_{cm}+y_s)\right\}
\right)
\end{eqnarray}
where $a$, $y_s$ and $w_s$ are parameters of the fit. The form
(\ref{2-sources-fit}) is a sum of two thermal sources shifted by $\pm
y_s$ from the midrapidity. The width $w_s$ of the sources can be
interpreted as $w_s=$ (temperature)/(transverse mass), if we assume
that collective velocities in the sources have no spread with respect
to the  source rapidities $\pm y_s$.

The above fit has been done by the least-squares method. 
Data were fitted in the rapidity range $|y-y_{cm}|/y_{cm}<0.7$. 
The choice of this range is dictated by the data. As a rule, the data
are available in this rapidity range, sometimes the data range is even
more narrow (80$A$ GeV and new data at 158$A$ GeV \cite{NA49-09}). 
The above constraint is imposed in order to treat different data in
approximately the same rapidity range. 
Another reason for this cut is 
that the rapidity range should not be too wide in order to
exclude contribution of could spectators.
%
%
I keep the old data at 158$A$ GeV 
\cite{NA49-1} in the analysis because these are known in a wider rapidity range 
as compared with the new ones \cite{NA49-09}. A narrow rapidity range results 
in large error bars of the fit. 
To evaluate errors of the fit parameters, I estimated the errors
produced by the least-squares method, as well as performed fits in
different the rapidity ranges: $|y-y_{cm}|/y_{cm}<0.5$ and 
$|y-y_{cm}|/y_{cm}<0.7$, where it is appropriate. 
The error bars present largest uncertainties among mentioned above. 

Similar fit was applied to the calculated distributions. 
Since experimental data at AGS and RHIC energies were taken 
from Au+Au collisions  
while at SPS the Pb+Pb collisions were studied, the calculations were performed 
respectively for Au+Au ($b=$ 2 fm) and Pb+Pb ($b=$ 2.4 fm) central collisions. 
In fact, at the same incident energy the computed results for 
Pb+Pb collisions at $b=2.4$ fm are very close to those for Au+Au at $b=2$ fm.
Therefore, the related irregularity of the energy dependence of the fit parameters
is negligible. 
Similarly to the experimental data, 
the fit of the computed results was also performed in two ranges,  
$|y-y_{cm}|/y_{cm}<0.7$ and $|y-y_{cm}|/y_{cm}<0.5$, 
in order to estimate uncertainty associated with 
varition of this range. 
This uncertainty turned out to be a dominant one 
in the case of computed data. 
Therefore, in Figs. 
\ref{fig4aa} and 
\ref{fig4a} results of the fit of 
computed spectra are presented by shaded areas with borders corresponding 
to the fit ranges $|y-y_{cm}|/y_{cm}<0.7$ and $|y-y_{cm}|/y_{cm}<0.5$.

Parameters  $y_s$ and $w_s$ deduced from
the fit of experimental data exhibits no significant irregularities in their energy dependence: 
they monotonously
rise with the energy within the error bars of the fit, see Fig. \ref{fig4aa}. 
Grey bands in all the panels of Fig. \ref{fig4aa} are drawn to guide an eye. They 
indicate the areas: between $0.25\, y_{cm}$ and $0.35\, y_{cm}$ for 
$w_s$ (lower  raw of panels), and between $0.4\, y_{cm}$ and $0.5\, y_{cm}$ for $y_s$ (upper raw of panels), 
where the major part of experimental points are located. 
In particular, similar (but based on 
different, double-gaussian fit) analysis of Ref. \cite{MehtarTani:2011uq} 
also shows absence of any spectacular irregularities in exitation functions of these parameters. 
This is more so in view of the fact that the  analysis of Ref. \cite{MehtarTani:2011uq} 
was performed only at $E_{lab} \geq$ 20$A$  GeV. 
The parameters deduced from the fit of  distributions computed within hadronic and crossover 
scenarios also manifest quite monotonous  behavior with incident energy. 
At the same time, the results of 2-phase scenario exhibit certain, however not strong, irregularity.

The representation in terms of $y_s$ and $w_s$ is not quite spectacular. These parameters are  
interrelated to some extent. They produce a similar effect on the rapidity distribution, especially 
if it is fitted in a narrow rapidity range. Therefore, it is desirable to find a single quantity 
which characterizes the shape of the rapidity distribution in the midrapidity range. 
Such a parameter is 
a reduced curvature of the spectrum in the
midrapidity defined as follows 
\begin{eqnarray}
\label{Cy} 
C_y &=& 
\left(y_{cm}^3\frac{d^3N}{dy^3}\right)_{y=y_{cm}}
\big/ \left(y_{cm}\frac{dN}{dy}\right)_{y=y_{cm}}
\cr
&=&  
(y_{cm}/w_s)^2 \left(
\sinh^2 y_s -w_s \cosh y_s 
\right). 
\end{eqnarray}
The factor $1/\left(y_{cm}dN/dy\right)_{y=y_{cm}}$
is introduced in order to get rid of overall normalization of the
spectrum, i.e. of the $a$ parameter in terms of fit
(\ref{2-sources-fit}). The second part of Eq. (\ref{Cy}) presents 
this curvature in terms of parameters of fit (\ref{2-sources-fit}).
Thus, the reduced curvature, $C_y$, and the midrapidity value of the distribution 
are two independent quantities quantifying the 
the spectrum in the midrapidity range. Excitation functions of $C_y$
deduced both from 
experimental data and results computed  
with different EoS's are displayed in Fig. \ref{fig4a}. 
Fig. \ref{fig4a} demonstrates how the distribution shape evolves from a convex 
form at low incident energies 
to a concave form at high energies. 

The irregularity in data 
is distinctly seen here 
as a zigzag irregularity in the energy dependence of $C_y$. 
Of course, this is only a hint to irregularity since this zigzag is formed only due to 
preliminary data of the NA49 collaboration. 
A remarkable observation is that 
the $C_y$  energy dependence in the first-order-transition
scenario manifests qualitatively the same zigzag irregularity 
(middle panel of Fig. \ref{fig4a}), as
that in the data fit, while the hadronic  scenario produces purely monotonous 
behaviour. 
The crossover EoS represents a very smooth  transition, as mentioned above. Therefore, 
it is not surprising that it produces only a weak wiggle in $C_y$.


As it was explained in detail in Ref. \cite{Ivanov:2012bh}, 
 the ``peak-dip-peak-dip'' irregularity is very natural in a system 
undergoing a 
phase or crossover transition. 
First, it is associated with the softest point 
of a EoS \cite{Hung:1994eq}. Therefore, the irregularity is weaker 
in the crossover scenario than in the first-order-transition one. Indeed, 
the softest points in the crossover EoS is less 
pronounced than in the first-order-transition one \cite{Nikonov:1998dg}. 
There is no softest point in the hadronic EoS and 
hence there is no irregularity.

The second reason of this irregularity is a change in the nonequilibrium regime. 
The 3FD model takes into account the leading nonequilibrium of the nuclear 
collision associated with a finite stopping power of the nuclear matter. 
It simulates the 
finite stopping power by means of friction between three fluids. Naturally, 
this friction changes when  
deconfinement  
happens. 
In the case of the crossover scenario this change in the friction is very 
smooth. Therefore, it does not contribute to the irregularity. 
At the same time this change in the friction enhances the irregularity in 
the first-order-transition scenario. 
As it was demonstrated in 
Ref. \cite{Ivanov:2010cu}, if the same friction is used in both phases, the
reduced curvature calculated with the 2-phase EoS exhibits only a weak wiggle
  in $C_y$  with considerably smaller amplitude  
as compared with zigzag in actual calculations with different frictions 
in different phases. These different frictions appear quite naturally 
in the 3FD model. The hadronic friction was estimated in Ref. \cite{Sat90} and works 
well at lower AGS energies. Therefore, there are no reasons to modify it. 
The partonic friction, while not microscopically estimated, is fitted to reproduce 
data at high incident energies. This is a reason to believe that it is a proper 
choice.

It is important to emphasize that the ``peak-dip-peak-dip'' irregularity 
is a signal from the hot and dense stage of the nuclear collision, 
rather than from the freeze-out stage as the most part of signal are.


\section{Summary}
\label{Summary}

Proton and net-proton rapidity distributions in collisions of heavy nuclei 
Au+Au (at AGS and RHIC energies) and Pb+Pb (at SPS energies) were analyzed
in a wide range of incident energies
 2.7 GeV  $\le \sqrt{s_{NN}}\le$ 39 GeV in terms of the 
center-of-mass energy per nucleon pair. The analysis was done 
within a model of the three-fluid 
dynamics  \cite{3FD} employing three different equations of state: a purely hadronic EoS 
\cite{gasEOS}
and two versions of EoS involving the deconfinement 
 transition \cite{Toneev06}. These are an EoS with the first-order phase transition 
and that with a smooth crossover transition. 
The crossover transition constructed in Ref. \cite{Toneev06}
is very smooth. The hadronic fraction survives up to very high energy densities. 
In this respect, this version of the crossover EoS certainly contradicts results of the 
lattice QCD calculations, where a fast crossover, at least at zero chemical potential, 
was found. Therefore, a true EoS is somewhere in between  the crossover and 2-phase EoS's 
of Ref. \cite{Toneev06}.

Scenarios based on EoS's with deconfinement transitions have a theoretical advantage as compared to the 
purely hadronic one. 
In order to reproduce the baryon stopping at high incident energies,
the friction between counter-streaming fluids have to be enhanced 
within the hadronic scenario 
as compared to its estimate based on experimental
inclusive proton--proton cross sections  \cite{Sat90}.  
Though such enhancement is admissible in view of uncertainties 
of the estimated friction, the value of the enhancement looks too high. 
In scenarios with deconfinement there is no need to modify the hadronic friction. 
This can be considered as an indirect argument in favor of such scenarios.

It was found that predictions within the  first-order-transition scenario, 
i.e. with the  2-phase EoS, exhibit a ``peak-dip-peak-dip'' irregularity in the 
incident energy dependence of the form of the net-proton rapidity distributions. 
At low energies, up to  $E_{lab}=$ 6$A$ GeV, rapidity distributions for central collisions 
have a peak at the midrapidity, similarly to results with other EoS's. 
Beginning from 8$A$ GeV, this peak turns into a dip at the midrapidity. 
Then again a peak is realized, starting from 
30$A$ GeV. 
With further energy rise ($E_{lab}>$ 40$A$ GeV) the midrapidity peak again transforms into a dip, 
which is already a normal behavior which takes place at arbitrary high energies.  
This behavior is in contrast with that for the hadronic scenario, 
where the form of distribution in central collisions gradually evolves from peak at the 
midrspidity (at $E_{lab}<$ 10$A$ GeV) to a dip (at $E_{lab}\gsim$ 10$A$ GeV).
The case of the crossover EoS is intermediate. Only  
a weak wiggle of the type of ``peak-dip-peak-dip'' is observed in the energy range of  
10$A$ GeV $\le E_{lab}\le$ 20$A$ GeV. 

The behavior the type of ``peak-dip-peak-dip'' in central collisions within
the 2-phase-EoS scenario is very robust 
with respect to variation of the model parameters in a 
wide range. It certainly disagrees with data at 8$A$ GeV, 10$A$ GeV and  40$A$ GeV energies. 
It also disagrees with data at 20$A$ GeV and  30$A$ GeV, which  
have preliminary status, and hence it is too early to draw any conclusions 
from comparison with them.

However, the experimental data also exhibit a trend of the ``peak-dip-peak-dip'' irregularity 
in the energy range 8$A$ GeV $\le E_{lab}\le$ 40$A$ GeV. Again this trend is based on 
preliminary data at energies of 20$A$ GeV and  30$A$ GeV. Therefore, 
updated experimental results at 20$A$ and 30$A$  GeV are badly needed 
to pin down the preferable EoS and to check the trend    of the 
 ``peak-dip-peak-dip'' behavior  in net-proton rapidity distributions. 
An irregularity in the baryon stopping is a signal of  
 deconfinement  
occurring in the compression stage of a nuclear collision. It is a combined effect 
of the softest point of a EoS and 
a change in the nonequilibrium regime from hadronic to partonic one.  
It is important to emphasize that this irregularity 
is a signal from the hot and dense stage of the nuclear collision.

An effective method to quantify the 
``peak-dip-peak-dip'' irregularity is the analysis of the distribution shape 
in terms of 
the reduced curvature of the spectrum in the
midrapidity $C_y$. In tems of $C_y$
this irregularity is distinctly seen  
as a zigzag irregularity in the energy dependence of $C_y$.

Comparison with available data, including those at RHIC energies,  
indicate certain preference of the crossover EoS, 
though  the crossover-EoS scenario in the presently used version does not perfectly reproduce the data either. 
Predictions of different scenarios for net-protons diverge to the largest extent
in the energy region  8$A$  GeV  $\leq E_{lab} \leq$ 40$A$  GeV. 
The preliminary data at 20$A$ and 30$A$  GeV disagree with any of the considered scenarios. 
At incident energies above the top SPS one the hadronic scenario 
certainly overestimates  the proton  midrapidity density, while deconfinement  
scenarios reasonably reproduce it.

\begin{figure}[bth]
\includegraphics[width=8.45cm]{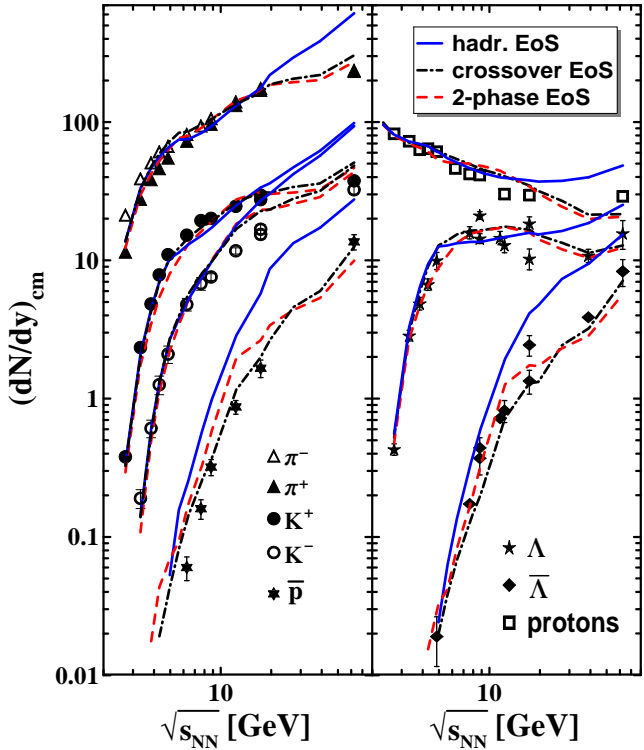}
 \caption{(Color online)
Midrapidity  densities  
 of various produced particles as functions of the incident center-of-mass energy
 of colliding nuclei predicted
 by 3FD calculations with  different EoS's. 
Experimental data are from compilation of Ref. \cite{Andronic:2005yp}
complemented by recent data from STAR collaboration \cite{Zhu:2012ph}
and  latest update of the  
compilation of NA49 numerical results \cite{Compilation-NA49,Blume:2011sb}. 
}  
\label{fig5}
\end{figure}

Anticipating results of subsequent papers, it should be mentioned that 
the 3FD simulations with same set of parameters described here 
also reproduce other observables, of course, with different degree 
of success depending on applied EoS. As an example, Fig.  \ref{fig5} 
demonstrates excitation functions of  midrapidity  values 
of various produced particles. 
As before, results for top 
calculated energy of $\sqrt{s_{NN}}=$ 62.4 GeV should be taken with care,  
since accurate computation is still unavailable for this energy.
The strangeness production at low incident
energies is overestimated within the 3FD model. 
\begin{figure}[th]
\includegraphics[width=5.0cm]{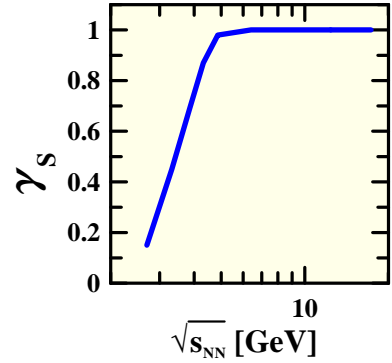}
 \caption{(Color online)
Strangeness suppression factor  as a function of the center-of-mass energy
 of colliding nuclei. 
}  
\label{fig6}
\end{figure}
This is not
surprising, since any EoS in the 3FD model
is based on the grand canonical ensemble. This shortcoming
can be easily cured by introduction of a phenomenological factor
$\gamma_S$ \cite{Koch:1986ud}, which
accounts for an additional strangeness 
suppression due to constraints of canonical ensemble. 
The midrapidity densities of strange particles, displayed in Fig.  \ref{fig5}, 
are multiplied by $\gamma_S$ factor, which in its turn is presented in  
Fig. \ref{fig6}. As seen, at $E_{\scr{lab}}>$ 10$A$ GeV there is no
need for additional strangeness suppression.

As seen from Fig. \ref{fig5}, the purely hadronic EoS certainly fails at high energies. 
A preferable EoS is the crossover one, similarly to that for proton rapidity 
distributions. This calculations, as well as those of other observables, will be discussed 
in subsequent papers.

\vspace*{3mm} {\bf Acknowledgements} \vspace*{2mm}

Fruitful discussions with 
I.N. Mishustin, L.M. Satarov 
and D.N. Voskresensky
are gratefully acknowledged. 
I am grateful to A.S. Khvorostukhin, V.V. Skokov,  and V.D. Toneev for providing 
me with the tabulated 2-phase and crossover EoS's. 
The calculations were performed at the computer cluster of GSI (Darmstadt). 
This work was supported by The Foundation for Internet Development (Moscow)
and also partially supported  by  
the Russian Ministry of Science and Education 
grant NS-215.2012.2.


\end{document}